\documentclass[aps,12pt,prb,final,notitlepage,oneside,onecolumn,nobibnotes,nofootinbib,noshowpacs,centertags]{revtex4}
\usepackage{graphicx}
\usepackage{dcolumn}
\usepackage{bm}

\begin{document}

\title{Nature of metal-nonmetal transition in metal-ammonia
solutions. II. From uniform metallic state to inhomogeneous
electronic microstructure}
\author{Gennady N.~Chuev$^{1}\footnote{e-mail addresses: genchuev@rambler.ru and pascal.quemerais@grenoble.cnrs.fr}$ and Pascal Qu\'{e}merais$^2$}
\date{\today}

\begin{abstract}
Applying semi-analytical models of nonideal plasma, we evaluate the
behavior of the metallic phase in metal-ammonia solutions (MAS).
This behavior is mainly controlled by the degenerate electron gas,
which remains stable down to 5 MPM due to high solvent
polarizability and strong dielectric screening of solvated ions.
Comparing the behavior of the metallic state with those of localized
solvated electrons, we have estimated the miscibility gap $\Delta n$
for various alkali metals and found $\Delta n$(Na)$> \Delta n($K$)$.
It is rather narrow in Rb-NH$_3$ and does not occur in
Cs-NH$_3$ solutions, which is in full agreement with the experiments.
The case of Li is discussed separately. The difference calculated in the
excess free energies of the metallic and nonmetallic phases is in
the order of $k_BT$, yielding a thermally fluctuating mixed state at
intermediate metal concentrations. It results in a continuous
metal-nonmetal (MNM) transition above the consolute point $T_c$ and
a phase separation below $T_c$.  We propose a criterion for the MNM
transition which may be attributed to the line of the maximum of
compressibility above $T_c$. This line crosses the spinodal one at the
critical temperature. Finally, we assert that a new electronic phase
similar to microemulsion should also arise between the
spinodal and the binodal lines.
\end{abstract}

\affiliation{$^{(1)}$ Institute of Theoretical and Experimental Biophysics, \\
Russian Academy of Science, Pushchino, Moscow Region, 142290, Russia \\
$^{(2)}$Institut N\'eel, CNRS, BP 166, 38042 Grenoble Cedex 9, France }

\maketitle

\newpage
\section{Introduction\label{sec1}}

Metal-ammonia solutions (MAS) are an example of quantum-classical
systems whose thermodynamic, dielectric, and optical properties are
controlled by electron-electron interactions. In our recent papers
\cite{comptes,1} we have explained that  these interactions are mainly
due to the dispersion attractions between solvated electrons at
metal concentrations in the range of 1-5 MPM.\cite{MPM} These attractions
result in two main effects: a phase separation in solutions of
light alkali metals below a critical temperature, and a dielectric
instability of the solution, which may be considered as the onset of
metallization. Considering the role of the induced dipolar
interactions between solvated electrons, we have restricted ourselves
 to the nonmetallic phase. \cite{1} We have already given an indication
\cite{comptes} that the metal-nonmetallic (MNM) transition, or
insulator-to-metal transition (IMT), are likely driven by a
Goldhammer-Herzfeld (GH) mechanism, \cite{goldhammer,herzfeld} i.e.
a polarization catastrophe which has its origin in the dispersion
interactions. We have also indicated that the correlation effects
\cite{comptes} and the solvent polarizability \cite{comment} may be
responsible for the unusual properties of the metallic phase and
lead to its instability at metal concentrations lower than 5 MPM. As
explained in Ref.\cite{comptes}, this seems to forbid the Mott-like
scenario for the MNM transition in MAS. That point reactivates the
old debates of Mott,\cite{mottbook,mottPRL,mottJPC} Jortner, and
Cohen \cite{jort,jort2,jo-co} on the nature of the MNM transition.
Mott assumed the phase separation to be a consequence of the
Mott-like transition, which is hidden by the phase separation below
the consolute temperature ($T<T_c$), and the MNM transition above
the temperature ($T>T_c$) should be of the Anderson type due to
disorder. On the contrary, Jortner and Cohen \cite{jo-co} described
the system above the consolute point as a microscopically
inhomogeneous regime in which the concentration fluctuates locally
one of the other well-defined values $M_0$ and $M_1$ ($M_0>M_1$),
and the MNM mechanism does not involve the Mott transition.
According to them, the concentration $M_1 \approx 2$ MPM corresponds
to the nonmetallic blue phase (consisting of solvated electrons as
described in Ref.\cite{comptes,1}), whereas the concentration $M_0
\approx 9$ MPM corresponds to the bronze metallic phase in which
electrons are delocalized and move freely. Using this hypothesis and
a semi-classical theory of percolation, Jortner and Cohen were able
to account for most transport properties observed in the
intermediate concentration regime above the consolute temperature.
\cite{jort,jort2,jo-co} However, in the conclusive part of their
paper,\cite{jo-co} they pointed out the four important unsolved
questions: 1) 'what is the origin of this microscopically
inhomogeneous state?'; 2) 'what is the origin of the phase
separation?';  3) is there any link between the two phenomena?; and
finally 4) 'what is the nature of the MNM transition?'.  We have
already partly answered the last question in our previous papers
\cite{comptes,1}, and we now think that we are able to answer the
three other ones, at least at a qualitative level at this stage of
our theory. This will be the subject of the discussion of the
present paper, but we are mainly facing a new theoretical situation.
Our results show that at low temperatures  an intermediate
concentration range exists in which both states, i.e., the
nonmetallic one (the blue phase) and the metallic one (the bronze
phase), are intrinsically unstable. \cite{comptes,1} Such a
theoretical possibility has been already suggested by one of us
\cite{quem1,quem2,quem3}. As we will discuss in this paper, we
believe that this is at the origin of the phase separation, and that
the existence of an inhomogeneous state above the consolute
temperature should additionally be clarified.

There are numerous experimental data on the electronic
\cite{thompson,burns,burns1,burns2,burns3} and structural
\cite{soper,soper01,soper2,soper3,burn3} properties of the metallic
state of MAS, but a theoretical treatment of concentrated MAS is
restricted to a few papers only. \cite{jort,jort1,jort2,Asch,K4,1a}
The difficulty of such a treatment is twofold: first, the electron
density in the metallic phase is still rather low so that strong
electronic correlation effects occur, and, secondly, the molecular
nature of the solvent as well as a the sufficient role of the
solvent polarizability complicate the theoretical study of these
correlations in the metallic phase. The aim of the present work is
to develop a statistical model treating the excess electrons in MAS
under conditions corresponding to the metallic phase. This phase
represents a three-component mixture consisting of delocalized
electrons, cations, and polarizable solvent molecules. In general, a
detailed information about electronic and thermodynamic properties
of the system may be obtained by quantum molecular simulations based
on the Car-Parrinello method. \cite{K2,deng,K3,K4} However, the
computational costs and sensitivity of such calculations to the
choice of the interaction potentials (an explicit account of
cations, polarizability of solvent and so on) restrict the
application of the method and cannot be used efficiently for a
complete understanding of the phase behavior of MAS. An alternative
way may be an employment of the integral equations theory based on
quantum hypernetted chain closure \cite{chihara}, but the current
status of such an approach is limited only to simple metals
\cite{anta} or electrons localized in polar liquids \cite{chuev,Bip}
and ionic liquids.\cite{chu1,chu2} Indeed, we do not know any
applications extended to molecular metallic fluids. In order to
avoid the complexity related  to numerical implementations, we
consider in the present paper that the metallic phase is well
described by an effective two-component plasma (TCP), where the role
of the solvent is restricted to its screening effects. We take into
account different frequency scales of this screening as we have done
for the nonmetallic phase, \cite{comptes,1} i.e. that the screening
of electron-electron interactions is due to the electronic
polarizability of solvent, whereas ion-ion interactions are
additionally screened by the permanent dipoles of the solvent
molecules. As a result, we use different dielectric constants to
treat the Coulomb interactions properly, i.e. a high-frequency
dielectric constant for the electron-electron interactions and a
low-frequency dielectric constant for ion-ion interactions
respectively. By using available analytical expressions for the
free-energy of the TCP \cite{ich,ich1}, we will derive the main
thermodynamic and dielectric characteristics of the metallic phase,
its pressure and chemical potential, and we will evaluate the locus
of the critical lines (spinodal and coexistence line) which
correspond to the instability of the metallic phase. Then, by
comparing the thermodynamic behavior of the metallic phase with that
of the nonmetallic phase,\cite{comptes,1} we will give evidence that
an inhomogeneous electronic microstructure must arise at
intermediate metal concentrations in MAS. Although the detailed
study of the inhomogeneous microstructure will be the subject of our
next paper, \cite{fut} we will assert in this work that the
thermodynamic and dielectric peculiarities of the inhomogeneous
electronic state govern the behavior of MAS in this concentration
range. Eventually, by treating the inhomogeneous microstructure
within the methods of simple liquids, we will propose a macroscopic
criterion for the MNM transition, and sketch the complete phase
diagram of MAS. Atomic units are used throughout.

\section{Model \label{sec2}}

\subsection{General outline.}

The high-density phases of MAS are composed of delocalized
interacting electrons scattered by ammonia molecules and  solvated
ions. As explained above, the solvent molecules are considered to
result in the screening of the ions and the electrons only. The
ion-ion interactions will be screened by the low frequency
dielectric constant $\epsilon_l$, while the electron-electron
interactions will be screened only by the high frequency dielectric
constant $\epsilon_h$. Next we treat the system as a TCP, the first
component being the interacting electron gas in a jellium, and the
second component being a classical plasma of ions in a jellium of
electrons. Finally, we add an interaction term which represents a
correction to the Madelung energy of the point ions in the jellium
of electrons, i.e. a pseudo-potential correction due to the
short-range scattering of the free electrons by valence core
electrons of the ions. Notice that these ions are themselves
solvated by ammonia molecules. This defines two different effective
radii for the ions. The first one is an effective classical radius
of the solvated ions, that we call $r_{vdW}$, which will be used to
account for the classical short-range ion-ion interactions through
the use of a packing factor. The second effective radius concerns
the electron-ion interactions and it is related to the core ion
radius, and we call it $r_{i}$. It includes the effect of the
solvation shell of a particular ion. For alkali metals such as Na,
K, Rb, Cs, the effective core radii are only slightly different from
those determined by the pseudo-potential method \cite{folias} for
non solvated ions. However, the case of Li$^+$ is special, since it
is well established \cite{soper01} that  owing to their small size,
the Li$^+$ ions occupy tetrahedral vacancies formed exactly by four
ammonia molecules. From this point of view,  Li$^+$ ions are
strongly bound to form complexes Li(NH$_3$)$_4^+$, which are
dominant ionic species in the solution. Consequently the core radius
of solvated Li$^+$ is substantially larger than in the case of
liquid Li metal.

Thus, our model contains four parameters $\epsilon_l$, $\epsilon_h$,
$r_{vdW}$, $r_{ion}$, in addition to the Wigner-Seitz parameter
$r_s= \hbar^2/me^2 (4 \pi n/3)^{1/3}$, related to the metal density
 (or the density of electrons, since the alkali metals are monovalent).
Finally, the excess free energy per electron of the metallic phase
is written as the sum of three terms:
\begin{equation}
f_m (n,T) = e_{DEG}(n) + f_{OCP}(n,T) + f_{ei}(n),
\end{equation}
where $e_{DEG}(n)$ is the free energy per electron of an interacting
electron gas with density $n$, and $f_{OCP}(n,T)$ is that of a
classical one-component plasma (OCP) of ions in a jellium of
electrons. Finally, the last term $f_{ei}(n)$ represents the
electron-ion interactions and will be treated in the framework of
the pseudo-potential theory \cite{folias}. Notice that Ashcroft
proposed a similar approach to the metallic phase,\cite{Asch} except
that he didn't used the different dielectric constants in his model.

\subsection{Free-energy of the interacting electron gas\label{sec21}}

The electron gas may be parameterized  by two dimensionless
parameters \cite{ich}: the electron coupling constant $\Gamma
_{e}=\beta e^{2}(4\pi n/3)^{1/3}/\epsilon _{h}$, and the reduced
temperature $\Theta =2m/\hbar ^{2}\beta (3\pi ^{2}n)^{2/3}$, where
$\beta $ is the inverse temperature. $m$ and $e$ are the electron
mass and charge, respectively. However one can easily check that
$\Gamma _{e}\sim 100\gg 1$ and $\Theta \sim 0.05\ll 1$ under
conditions corresponding to the metallic phase in MAS, and therefore the
electron gas is  degenerate and we can treat it at $T=0K$. Its free
energy may be parameterized by the single parameter, i.e. the
effective Wigner-Seitz radius defined as $r_s^*=r_s/ \epsilon_h$.
Applying the conventional methods for the electron gas, we have:
\begin{equation}
\left( \frac{\epsilon _{h}^{2}\hbar ^{2}}{me^{4}}\right) e_{DEG}(n)=\frac{1.105}{r_{s}^{\ast 2}}-%
\frac{0.458}{r_{s}^{\ast }}+e_{cor}(r_{s}^{\ast }).
\end{equation}
The first two terms are obtained by the usual Hartee-Fock
approximation, and the last term is the correlation energy (beyond
the Hartree-Fock approximation). In order to treat this correlation
energy $e_{cor}(r_s^*)$, which is a smooth function of $r_s^*$, we
have employed the result of the local density approximation
parameterized in Ref. \cite{Pedrew} by fitting quantum simulations:
\begin{equation}
e_{cor}(r_s^*)=\frac{\gamma _{0}}{1+\gamma _{1}r_{s}^{\ast }{}^{1/2}+\gamma _{2}r_{s}^{\ast }},
\end{equation}
where $\gamma _{0}=-0.1423$, $\gamma _{1}=1.0529$, and $\gamma
_{2}=0.3334$ are numerical parameters obtained from ref.
\cite{Pedrew}. As we noticed in Ref. \cite{comptes}, the degenerate
electron gas gives the main contribution to the total excess free
energy $f_m(n,T)$ of the metallic phase.

\subsection{The classical one-component plasma\label{sec22}}

The mean distance between ions exceeds 9 \AA\ at metal
concentrations of about $10$ MPM, so short-range details of the
ion-ion interactions can be ignored, and we may treat the ions as
charged hard-spheres in the uniform jellium background. Thus, the
excess free energy of ions is controlled by the two dimensionless
parameters: the dimensionless ion coupling constant $\Gamma
_{i}=\beta e^{2}(4\pi n/3)^{1/3}/\epsilon _{l}$, and the packing
factor $\eta = \pi n\sigma^{3}/6$ in which $\sigma=2r_{vdW}$ is the
classical hard-sphere ion diameter. Simple evaluations show that
the first parameter is of about 10 in the metallic Na-NH$_3$, and that
the second one does not exceed 0.05 under the same conditions.
Therefore the free energy $f_{OCP}$ per ion may be written as
\begin{equation}
\beta f_{OCP}=[\ln (n\Lambda_i^3)-1]+f_{hs}(\eta)+f_{C}(\Gamma_i),
\label{OCP1}
\end{equation}
where  the first term is the ideal contribution,
$\Lambda_i=(2\pi\beta/M_i)^{1/2}$ is the de Broglie length and
$M_i$ is the ion mass. The second term in (\ref{OCP1}) is the
hard-sphere contribution, and the last term  is the contribution
due to Coulomb interaction between ions. Analytical expressions for
these contributions are well-known. The hard-sphere part is
expressed as \cite{CS}:
\begin{equation}
\beta f_{hs}(\eta) = \frac{\eta (4-3\eta )}{(1-\eta )^{2}}.
\end{equation}
According to Ref.\cite{DWI}, in which simulation data were fitted for various
$\Gamma_i $ in the range $1<\Gamma_i <160$, the electrostatic
contribution can be approximately written as:
\begin{equation}
\label{fc}
\beta f_{C}(\Gamma_i)=-0.897\Gamma_i +3.620\Gamma_i ^{1/4}-0.758\Gamma_i ^{-1/4}-0.815\ln \Gamma_i -2.58.
\end{equation}
The first term in this expression represents the Madelung energy and
gives the most important contribution. The remaining terms are
temperature-dependent corrections due to thermal motion of the ions.

\subsection{Electron-ion interaction\label{sec23}}

The main difficulty is to evaluate $f_{ei}(n)$, because the bare
ion-electron interaction is modified in the solution by the
polarizability of the solvent. Moreover, the Madelung term  in eq.
(\ref{fc}) concerns the point ions. The correction to this
approximation is well treated by the pseudo-potential model for
simple metals \cite{folias}. In that point we follow the Ashcroft
approach \cite{Asch} and adapt the pseudo-potential model
\cite{folias} to our case. As a result,  we write the electron-ion
contribution (expressed in effective atomic units $me^{4}/\epsilon
_{h}^{2}\hbar ^{2}$) as:
\begin{equation}
\left( \frac{\hbar ^{2}\epsilon _{h}^{2}}{me^{4}}\right) f_{ei}(n)=\frac{3r^{*2}_i}{2r^{\ast 3}_s},
\end{equation}
where $r^*_i=r_i/\epsilon_{h}$ and the ion-core parameter $r_{i}$ is
related to the atomic number of the ion as discussed above.
Notice that in our previous paper, \cite{comptes} we took a slightly
different notation and wrote $f_{ei}(n)=a_in$, so that $a_i=6 \pi
r_i^2$ to make the exact link with the present paper.

\section{Thermodynamical properties}

Once we know the expression of the excess free energy $f_m$ per electron, we can deduce the excess  pressure $%
\Delta p$, the excess chemical potential $\Delta \mu $, and the excess compressibility $\kappa$  as determined by the usual formulas:
\begin{equation}
\Delta p=n^{2}\frac{\partial f_m}{\partial n},\qquad \Delta \mu =f_m+n\frac{%
\partial f_m}{\partial n}, \qquad \kappa^{-1} = n \frac{\partial \Delta p}{\partial n}.  \label{press}
\end{equation}
As discussed above, it may be checked that the degenerate electron
gas (DEG) gives the main contribution to $f_{m}(n,T)$ at low
temperatures, whereas the ionic contribution is only a correction.
Our TCP model of the metallic phase of MAS depends on temperature
and metal concentration and can be  characterized by the four
parameters: $r_{vdW}$, $r_i$, $\epsilon_l$, and $\epsilon_h$. As we
will see, the value of the ionic mass $M_i$ in (\ref{OCP1}) does not
change the values of the density at which the metallic state becomes
unstable.

Earlier on \cite{1}, we have determined the low-density spinodal line
$n_s(T)$ above which the solvated electrons become thermodynamically
unstable. Now we evaluate the high-density counterpart of the
spinodal line $n_{c2}(T)$ corresponding to the zero derivative of
the excess chemical potentials or the excess pressure:
\begin{equation}\partial \Delta\mu/\partial n=\partial \Delta p/\partial n=0.
\end{equation}
Since the main part of the excess free energy  in the metallic phase
is coming from the degenerate electron gas, the high-density
spinodal $n_{c2}(T)$ will depend weakly on  temperature.

For any second-order phase transition there are two curves: the
spinodal and the binodal ones, the last curve corresponds to the
liquid-liquid coexistence. As we explained in Ref. \cite{1}, the
low- and high-density spinodals are given by $n_s(T)$  and
$n_{c2}(T)$, respectively. In principle, to calculate the
coexistence curve, we need a complete knowledge of the excess free
energy $f(n,T)$ in the whole range of density, which is beyond the
current status of our theory. However, we may use the following
argument to estimate the high-density part of the coexistence curve:
the excess pressures of the low-density and high-density phases must be equal (this is the coexistence condition). But we know
 that the excess pressure of the low density phase is
roughly close to zero (more exactly it is in the order of $n
k_BT$),\cite{1} whereas the excess pressure of the metallic phase
strongly varies with $n$. Therefore, we assert that the coexistence curve
$n_p(T)$ at high density will be roughly given by the condition:
\begin{equation}
\Delta p(n_p)=0. \label{zero}
\end{equation}

 The critical concentration $n_{c2}(T)$ defines the
low-density limit of the absolute stability of the metallic phase,
whereas the critical concentration $n_{p}(T)$ is close to the line of
liquid-liquid coexistence. The metallic phase cannot exist below
$n_{c2}(T)$, and in the range $n_{c2}(T)\le n\le n_p(T)$ the
solution is not homogeneous (the pressure becomes negative) and
consists of domains of the metallic phase together with nonmetallic
islands being nuclei of the low density phase, i.e. the solvated
electron phase. The uniform metallic phase remains stable for densities
$n>n_p(T)$.

\section{The use of experimental data to fit the model parameters \label{sec22}}

Our input phenomenological parameters are $r_{i}$, $r_{vdW}$,
$\epsilon_h$, and $\epsilon_l$. For the pure ammonia solvent, we
take $\epsilon_\infty=1.756$, $\epsilon _{s}(T=-70^{0}$C$)=25$, and
$\partial \epsilon _{s}/\partial T=-0.1\ $K$^{-1}$.\cite{epsT}  We
use  the data on the ammonia density, \cite{rhoT} namely,
$n_{NH_{3}}=0.0255$\AA $^{-3}$ at $T=-70^{0}$C. In order to take
into account phenomenologically the influence of  free electrons on
the high-frequency dielectric constant, we apply:
\begin{equation}
\epsilon_h(n)=\epsilon _{\infty }-Ac, \label{epsh1}
\end{equation}
where $c=n/(n+n_{NH_3})$ is the relative fraction of the metal in
the solution, and $A$ is a phenomenological parameter. In general,
the low-frequency dielectric constant $\epsilon_l$ may deviate also
from the bulk value $\epsilon _s$ due to saturation of orientational
polarization in the vicinity of ions. Simple estimates show that the
low-frequency constant can decrease down to 9. \cite{lik1}
Therefore, different values for $\epsilon_l$ could also be tested.
The core-radii and van-der-Waals radii can be extracted from the
literature, which we do in our first parameterization (model 1).
However, as we discussed above, this does not take properly into
account the fact that the ions are solvated. Then we propose a
second procedure to evaluate these radii, that we call model 2.

\textbf{{Model 1.}} The ion-core radii $r_i$ are obtained from the
data on simple metals. \cite{folias} Although we do not find
suitable estimates of van-der-Waals radii $r_{vdW}$ for ions
solvated in ammonia, we have used those of hydrated ions, assuming
that the solvation properties of water are not so different from
those of ammonia. We take the values for $r_{vdW}$ from Ref.
\cite{jorg}. That gives the values reported in Table 1 (at the
lines: model 1). Finally, in this first model, we apply the
phenomenological parameter $A=1-\epsilon_\infty=0.756$, that takes
into account the volume fraction occupied by the metal as in Ref.
\cite{thompson1}.

\textbf{{Model 2.}} For this model, the phenomenological constant
$A$ is used as a fitting parameter of experimental data on plasmon
excitation measured  in Li-NH$_3$. \cite{burns1,1a} It is
experimentally found $\epsilon_h=1.44$ at  20 MPM so that $A=1.58$.
As discussed above, the case of Li raises a question, since the core
radii obtained from Ref. \cite{folias} do not take into account
the solvation shell of the ion. We use $r_i$ as a parameter to fit
the experimental data on the phase coexistence, which is known for
 Na-NH$_3$, Li-NH$_3$, and K-NH$_3$ solutions. \cite{rob} We fit
these data by (\ref{zero}), and obtain the values for the effective
ion-core radius. A similar method is often used to fit ion-core
radii for simple liquid metals. \cite{Ach1} Comparing $r_i$ to
$r_{vdW}$ for model 1, we find a proportionality constant of about
$3/2$  for Na-NH$_3$. Then we use the same proportionality constant
to deduce $r_{vdW}$ from $r_i$ for Li, K, and Rb, which gives the
values reported in Table 1 (at lines: model 2).

\bigskip
\noindent Table 1. Parameters of the metal ions and the calculated critical densities for models 1 and 2.\\

\begin{tabular}{|l|l|l|l|l|l|l|l|l|}
\hline & \multicolumn{4}{c|}{model 1} & \multicolumn{4}{c|}{model 2}
\\ \hline  ion & $r_{i}$(\AA ) \cite{folias} & $r_{vdW}$(\AA )
\cite{jorg} & $n_{s}(MPM)
$ & $n_{c2}(MPM)$ & $r_{i}$(\AA ) & $r_{vdW}$(\AA ) & $n_{s}(MPM)$ & $%
n_{c2}(MPM)$ \\ \hline  Li & 0.761 & 1.33 & 1.1 & 7.1 & 1.25 & 1.88
& 1.72 & 3.8 \\  \hline   Na & 0.939 & 1.68 & 1.32 & 4.92 & 1.12 &
1.68 & 1.32 & 5.2 \\ \hline   K & 1.234 & 2.02 & 2.04 & 3.0 & 1.31 &
1.97 & 1.96 & 3.4 \\ \hline  Rb & 1.323 & 2.16 & 2.76 & 2.65 &1.33
&2.0 &1.96 & 3.2\\ \hline Cs & 1.449 & 2.54 & 5.0 & 2.15 &  &  &  &
\\ \hline
\end{tabular}

\section{Analysis\label{sec3}}
\subsection{Pure metallic phase\label{sec31}}

Using the formulas described above, we have calculated the excess
chemical potential and the excess pressure for the metallic phase of
Na-NH$_3$ solutions. Figure 1 shows the concentration dependencies
of these thermodynamic characteristics at various temperatures for
model 1. As can be seen, the chemical potential has the minimum at
$n_{c2}\approx 5$ MPM and the pressure is negative below $n_{cp}=9$
MPM at low temperatures. Therefore, the metallic phase is not stable
at densities lower than $n_{c2}$. This instability results from
Coulomb correlations between delocalized electrons, because the main
contribution to this compressibility is due to the DEG. It is well
known (see f.e. \cite{ich2}) that the compressibility of a pure DEG
in jellium (without polarizable medium), becomes negative at $r_s
\ge 5.24$ and its pressure is negative for $r_s \ge 4.18$, which
would correspond to $n_{c2}=30$ MPM and $n_{p}=46$ MPM,
respectively. These estimates are far from the experimental values
in MAS, which clearly indicates that the high solvent polarizability
and strong dielectric screening of ions stabilize the DEG in MAS to
lower metal concentrations. The temperature influence on the
stability is due to the ionic contribution, and it is as small as it
should be for a quantum phase transition. The critical concentration
$n_{c2}$ varies only from 5.5 to 4.5 MPM as the temperature changes
by $100^0$ C. The critical concentrations obtained are sensitive to
the choice of the input parameters $r_i$, $\epsilon_s$, and $A$.
Figures 2-4 show the variations of these critical concentrations
with respect to these parameters. A decreasing contribution of the
solvent polarizability tends mainly to destabilize the metallic
phase, since $n_{c2}\propto \epsilon_h^{-3}(n)$, and an increasing
ion-core radius promotes the phase stability at lower densities.

In our theory, we can estimate  the miscibility gap by considering
the range between the critical lines obtained at low density
$n_s(T)$ (our previous paper \cite{1}) and the present calculations
of $n_{c2}(T)$. Both the lines $n_{s}(T)$ and $n_{c2}(T)$ give the
locus of the low- and high-density spinodals, while the range
$n_{s}\le n\le n_{c2}$ correspond to the miscibility gap. The
results calculated at $T=-70^0$C are listed in Table 1. We find that
the gap decreases monotonically as the ion size rises as it happens
for simple metals and the gap disappears for heavy ions like Cs,
which is experimentally confirmed. \cite{thompson} Although these
qualitative predictions provide a correct trend for the miscibility
gap versus the ion size, the estimates based on the ion-core
parameter derived from the theory of simple metals\cite{folias} (our
model 1) do not yield accurate evaluations of the phase separation
range in case of Li-NH$_3$. The reason of this discrepancy is a
peculiarity of local microstructure around the solvated ions, as we
discussed above. Namely, the size of the Li ion is small, and the
coordination number for the solvated Li ions is small too in
comparison to other ions. As a result, the delocalized electrons
scatter on the solvated Li-NH$_3$ complexes rather than on the cores
of Li ions, and the effective ion-core size deviates sufficiently
from the value obtained for simple metals. This drawback disappears
when we apply model 2 and use the data on $r_i$ fitting the
zero-pressure condition (\ref{zero}) at the coexistence line. The
critical concentrations $n_{s}$ and $n_{c2}$ calculated in this
manner correspond much better to the experimental data (see Table
1); in particular, they yield the correct trend in the range $\Delta
n=n_{c2}-n_{s}$ of the phase separation, namely, $\Delta n$(Na)$>
\Delta n$(Li)$> \Delta n($K$)$. Both models 1 and 2  also indicate
the absence of the phase separation in Cs-NH$_3$ solutions, while
our results for  Rb-NH$_3$ solutions exhibit a possibility a phase
separation for model 2 and absence for model 1. The experimental
evidence of such separation is not clear. There is no visible phase
separation like in the Li-NH$_3$ solutions, \cite{rob} although the
measurements of conductivity \cite{thompson1} and X-ray scattering
\cite{Rb} indicate large fluctuations at $n_c\approx$ 4 MPM and at
temperatures close to $T_c=197^0$K. Those fluctuations can be
interpreted as an indirect evidence of a phase separation for
Rb-NH$_3$ solutions in this range. \cite{thompson1,Rb}

 We have also evaluated the
compressibility $\kappa$ as a function of the metal concentration
$n$ in the framework of  model 2. The inverse compressibility
$\kappa^{-1}$ calculated is shown in Fig. 5 together with the
experimental data derived from the measurements of plasmon
excitations \cite{burns2} obtained for the metallic Na-NH$_3$ and
Li-NH$_3$ solutions above the consolute point $T_c$. As can be seen,
there is a qualitative agreement between the calculated and the
experimental data. Both indicate that the inverse compressibility
decreases as the metal concentration decreases. However, the
theoretical evaluation gives a much higher compressibility than the
experimental measurements at intermediate densities. The deviations
are more pronounced as the concentration decreases. In our opinion,
it is an indirect evidence of the formation of an inhomogeneous
electronic state consisting of a microscopic mixture of delocalized
and localized electrons, as will be discussed in Sec. \ref{sec32}.

Finally, with the use of the calculated compressibility we can
evaluate the low-frequency dielectric response of the metallic state
at small wave-vectors ($k \rightarrow 0$), since  we have the
limiting relation for this function, as it takes place for the usual
electron gas:
\begin{equation}
\varepsilon (\omega \rightarrow 0,k\rightarrow 0)=\varepsilon _{h}(n)+\frac{%
q_{TF}^{2}}{k^{2}}\frac{\kappa _{free}}{\kappa },
\end{equation}
where $q_{TF}$ is the Thomas-Fermi screening wave vector equal to
$2(3n/\pi )^{1/6}/a_{0}$ ($a_{0}$ being the Bohr radius), and
$\kappa _{free}=(3/\pi ^{4}n)^{1/3}/n$ is the compressibility of the
ideal Fermi gas. The function $\varepsilon ^{-1}(k)=1/\varepsilon
(\omega =0,k\rightarrow 0)$ is depicted in Fig. 6 at various metal
concentrations. This function becomes negative at concentrations
below $n_{c2}(T)$, which confirms that the region of phase
separation is indeed a region where the static dielectric constant
must be negative. \cite{quem2,quem3}

The experimental phase diagram of sodium-ammonia solutions
\cite{crauss,CHIEUX,rob} together with the different calculated
critical lines corresponding to various instabilities of the
nonmetallic and the metallic phases are depicted in Fig. 7. The
low-density  and the high-density spinodals correspond to the
critical lines $n_s(T)$ and $n_{c2}(T)$, respectively, and the line
$n_{c1}(T)$ of polarization catastrophe (see paper \cite{1}) gives
the onset of the MNM transition. Although our calculations of
critical lines roughly correspond to the experimental situation, the
experimental coexistence line coincides with the theoretical
zero-pressure line of the metallic phase only at low temperatures
and deviates significantly from it as the temperature rises. The
situation is similar for the calculated low-density and high-density
parts of the spinodal curve, they do not cross as they should at the
consolute point $T=T_c$. Therefore, at this stage of the theory, our
model of the uniform metallic state and the homogeneous phase of
solvated electrons phase cannot explain all the peculiarities of the
phase behavior of MAS, although it indicates the main features of
this behavior, namely, the existence of a miscibility gap giving
rise to a phase separation.

\subsection{Thermally fluctuating inhomogeneous state\label{sec32}}

To qualitatively understand what happens in the intermediate region,
we have calculated the difference  $f_{nm}-f_m$ between the
excess free energies of the nonmetallic and the metallic phases.
Following our previous study, \cite{1} the free energy per electron
in the nonmetallic state can be expressed as:
\begin{eqnarray}
&&f_{nm}(n) =\frac{9}{8r_{e}^2}-[\frac{1}{\varepsilon _{\infty }}%
- \frac{1}{\varepsilon _{s}}]\frac{1}{2r_{e}} + 4\pi C_{S}\lambda r_{e}^{2}+ \frac{4\pi C_{V}n_{NH_3}}{3\beta }r_{e}^{3}+  \\
&+&2\beta^{-1} \left[ \ln
(n\Lambda_i^{3/2}\Lambda_e^{3/2})-1+\frac{\eta_{nm} (4-3\eta_{nm}
)}{(1-\eta_{nm} )^{2}}\right]-\frac{2}{\epsilon _{s}[\sigma +\gamma
^{-1}]}+\frac{\gamma ^{3}}{3\pi n\beta} -\frac{C_{\alpha} n
r_{e}^{3}}{\varepsilon _{\infty }^{2}}, \nonumber \label{fhs}
\end{eqnarray}
where $C_V$, $C_{S}$, and $C_{\alpha}$ are numerical coefficients
that we have already discussed in Ref. \cite{1} and related to the
temperature- and concentration- dependencies of the radius
$r_e(n,T)$ of the solvated electrons,
$\Lambda_e=(2\pi\beta/M_e)^{1/2}$ is the de Broglie length for
localized electrons, $M_e$ is their effective (classical) mass,
$\lambda$ is the surface tension, and $\eta_{nm}=\pi
n(r_e+r_{vDW})^{3}/3$ is an effective packing factor, that we take
equal to the mean value between the solvated electron and ion
diameters. Eventually, $\gamma =([1+(32\pi
n/\epsilon_{s})^{1/2}(r_e+r_{vDW})]^{1/2}-1)/2(r_e+r_{vDW})$ is the
inverse screening length. The first row of the expression yields the
solvation free energy of noninteracting  electrons, and the second
row results from electron-electron interactions and includes
short-range, electrostatic, and dispersion contributions,
respectively. All these contributions are only corrections to the
free-energy of the electron solvation energy. \cite{1}

 To calculate $f_{nm}$, we have used our model described
in Ref. \cite{1}, namely: $C_{k}=1.5$, $C_{r}\approx 1.25$,
$C_{S}=1$ and $ C_{V}=1.75$, $\lambda=40$ dyn/cm, and we apply
$C_{\alpha}=0.14$ and $\Lambda_e=0.27$\AA. The difference $\delta f
(n)=f_{nm}(n)-f_m(n)$ is shown in Fig. 8a. The main point to
underline is that their difference remains in the order of only a
few $k_B T$ along the range of concentration 1-10 MPM, although the
values of the free energies of the metallic and non-metallic state
are both in the order of -0.7 eV. We should emphasize that this is
not by chance. By carefully examining (and simplifying) the energy
of both the metallic and non-metallic states, we can indicate that
the main part of the energies coming from the electronic part is
roughly given by:
\begin{eqnarray}
f_{nm} \approx \frac{1.12}{r_e^2}- \left(
\frac{1}{\epsilon_{\infty}}- \frac{1}{\epsilon_{s}} \right)
\frac{0.5}{r_e},  \qquad f_m  \approx \frac{1.1}{r_s^2} -
\frac{0.46}{\epsilon_{\infty}
r_s}+\frac{e_{cor}(r_s/\epsilon_{\infty})}{\epsilon^2_{\infty}},
\end{eqnarray}
where $f_m$ stands for the metallic phase, and $f_{nm}$ for the
nonmetallic one. That clearly indicates a crossing between the two
energies when $r_s$ tends to  $ r_e$, which is in quantitative
agreement with the results shown in Fig. 8a.

Consequently, the thermal fluctuations between these electronic
states must play a central role at intermediate densities, in
particular to allow the closure of the miscibility gap at the
critical temperature $ T_c$ by a mixing entropy effect. The detailed
calculations of the complete phase diagram, including the
calculation of $T_c$ will be reported in our future paper.
\cite{fut} However, at a qualitative level, we may say that above
the critical temperature, the system should be a microscopic mixture
of both states, which is highly thermally fluctuating and roughly
described by the relative fractions
$n_m=(1+\exp[\beta(f_{m}(n)-f_{nm}(n))])^{-1}$ and $1-n_m$ of
electrons in the metallic and nonmetallic state respectively, as a
function of the metal concentration. The calculated values of $n_m$
are also shown in Fig. 8b for  Na-NH$_3$ solutions at $T=240^0$ K.
They indicate that our initial criterion of metallization, i.e. the
polarization catastrophe above $n_{c1}(T)$ \cite{comptes,1} is
spread out above $T_c$ and becomes a progressive MNM transition,
which should take place around the density $n_{c1}(T)$. Another
peculiarity of the MNM transition is the compressibility of the
mixed state consisting of localized solvated electrons as well as
delocalized free electrons. The line of maximum of compressibility
(or minimum of the inverse compressibility) may well characterize
the MNM transition. This is sketched on Fig. 9 by the total inverse
compressibility of the mixture. The total compressibility is well
given by the sum of the inverse compressibility for each species
(solvated electrons and delocalized ones) at the respective
densities $1-n_m$ and $n_m$. At the same time, the dielectric
constant remains finite at the line of the minimum inverse
compressibility, although it rises sharply in the nonmetallic phase
close to this line, as discussed in our previous paper\cite{1}. It
is worth noting that the mixed state qualitatively described here,
is of a new kind in condensed matter (to our knowledge), and really
results from the competition, driven by strong thermal fluctuations,
between the delocalized and the self-trapped quantum states.

The situation is more simple below $T_c$.  There is  a finite range
of densities between the spinodal and binodal curves at each side of
the phase diagram, where the minority phase may nucleate in the
majority phase to form some kind of stable electronic microemulsion.
This microemulsion phase would be characterized by a large variety
of aggregates  as reported in numerical simulations \cite{K4}. At
the same time, the solution is phase separated below the spinodal
line to prevent negative compressibility. The Fig. 10 summarizes all
these effects. The MNM transition may be assigned to the line of the
minimum inverse compressibility which crosses the spinodal and
coexistence lines at the critical point $n_c(T_c)$. A more detailed
theory of all these effects will be proposed in our future publication
as well as more precise calculations of the spinodal
and coexistence lines \cite{fut}. We have only qualitatively discussed in this paragraph
the main features of the phase diagram resulting from our model.

\section{Conclusion}

Thus, using the methods of non ideal plasma, we have evaluated the
behavior of the pure metallic phase in MAS. We have asserted that
this behavior is mainly controlled by the state of degenerate
electron gas. Due to high polarizability of ammonia and high
dielectric screening of ions solvated in MAS, the gas remains stable
up to values of $r_s$ of about 11, which is quite different from
simple metals where available values of $r_s$ do not exceed 6.  The
pure metallic phase is unstable at metal concentrations lower than 5
MPM. Comparing this critical concentration $n_{c2}$ with the
critical concentration $n_s(T)$ corresponding to the van-der-Waals
instability of solvated electrons, we have evaluated the range
$\Delta n$ of the phase separation and found it to be governed by
the ion-core size, the range decreasing as the ion-core size rises.
By evaluating the ion-core size with the use of the zero-pressure
condition applied to the coexistence line, we have obtained the
correct trend in the range of the phase separation, namely, $\Delta
n$(Na)$> \Delta n$(Li)$> \Delta n($K$)$. The phase separation does
not occur in Cs-NH$_3$ solutions and is rather narrow in the case of
Rb-NH$_3$.

Comparing the calculated free energy of pure metallic and
nonmetallic phases, we have asserted the difference between the
energies to be small due to minor differences in the energy of
polaron formation and that of electron gas at the stability
boundary. This leads to a strong influence of thermal fluctuations on
the nature of the MNM transition, which becomes continuous above the
critical temperature $T_c$. A thermally fluctuating inhomogeneous
electronic mixture arises. It modifies substantially the
criterion for the MNM transition, because this transition should be
attributed to the line of the minimum inverse compressibility or to
the maximum scattering factor. This line is quite close to our previous
estimates $n_{c1}(T)$ corresponding to the polarization catastrophe
and divergent dielectric constant, \cite{1} although the dielectric
constant remains finite at temperatures exceeding $T_c$.

At the same time, the dielectric constant and the compressibility
must be negative below the spinodal line. To avoid such negativity,
the system macroscopically separates into the metallic and
nonmetallic phases below the spinodal line. A new electronic state
arises in the range restricted by the spinodal and the binodal
lines. This state can be characterized as an electronic
microemulsion. The locus of the critical point $n_c$ and the
spinodal line is determined by the thermodynamic behavior of
microemulsion and will be the subject of our next study. \cite{fut}
Concerning the criterion for the MNM transition, we should note that
it is quite different from the one for usual semiconductors
\cite{mott} and indicates a sufficient role of thermal fluctuations
in the mechanism of MNM transitions. On the other hand, the line of
the minimum inverse compressibility always exists for the second
order transitions in simple \cite{srk} and molecular \cite{stanley}
liquids as well as for self-assembled networks. \cite{safran} This
line crosses the spinodal line at the critical point. That is the
reason why the MNM transition couples with the phase separation in
MAS.

\begin{acknowledgements}
GNCh thanks the Russian Foundation for Basic  research for partial support of this work.
\end{acknowledgements}

\newpage

\begin{figure}[tbp]
\includegraphics{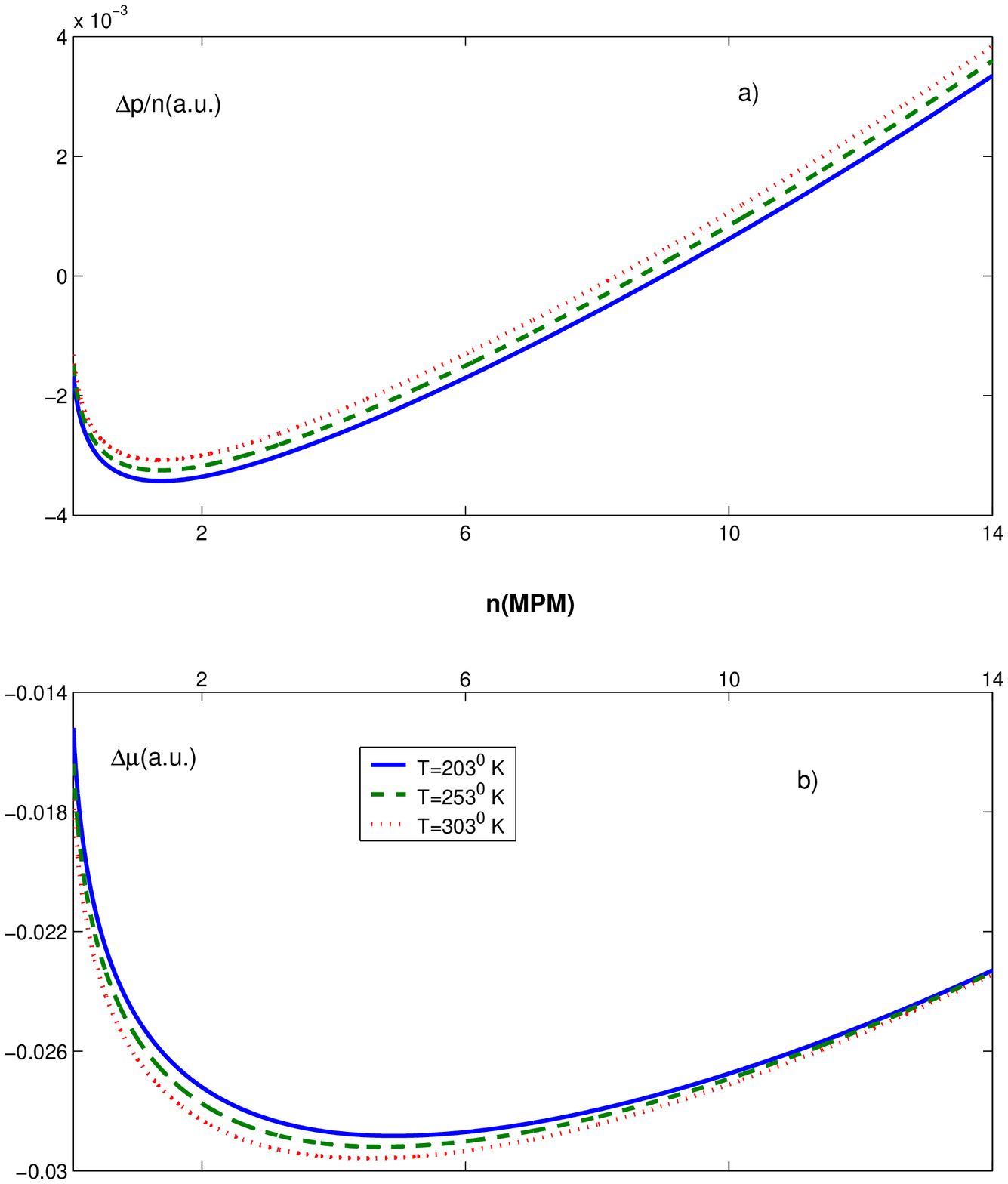}
\caption{ The  excess pressure $\Delta p(n,T)/n$ (a) and the excess
chemical potential $\Delta \mu (n)$ (b) versus metal concentration
$n$ at various temperatures for the metallic phase of Na-NH$_3$
solution (model 1). }\label{fig1}
\end{figure}

\begin{figure}[tbp]
\includegraphics{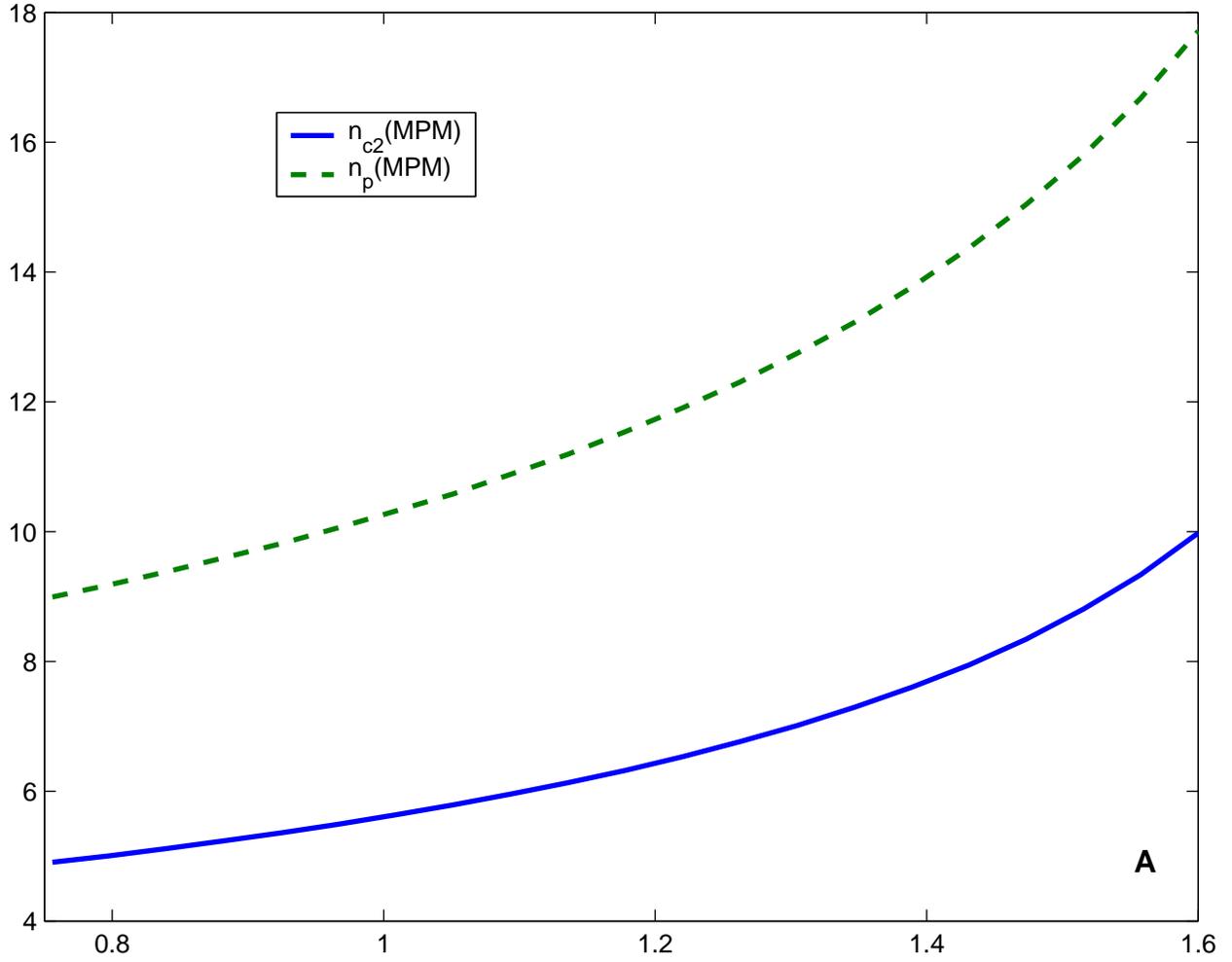}
\caption{Critical concentrations $n_{c2}$ and $n_p$ versus the
parameter $A$ for the metallic phase of Na-NH$_3$  at $T=-70^0$C
(model 1).} \label{fig2}
\end{figure}

\begin{figure}[tbp]
\includegraphics{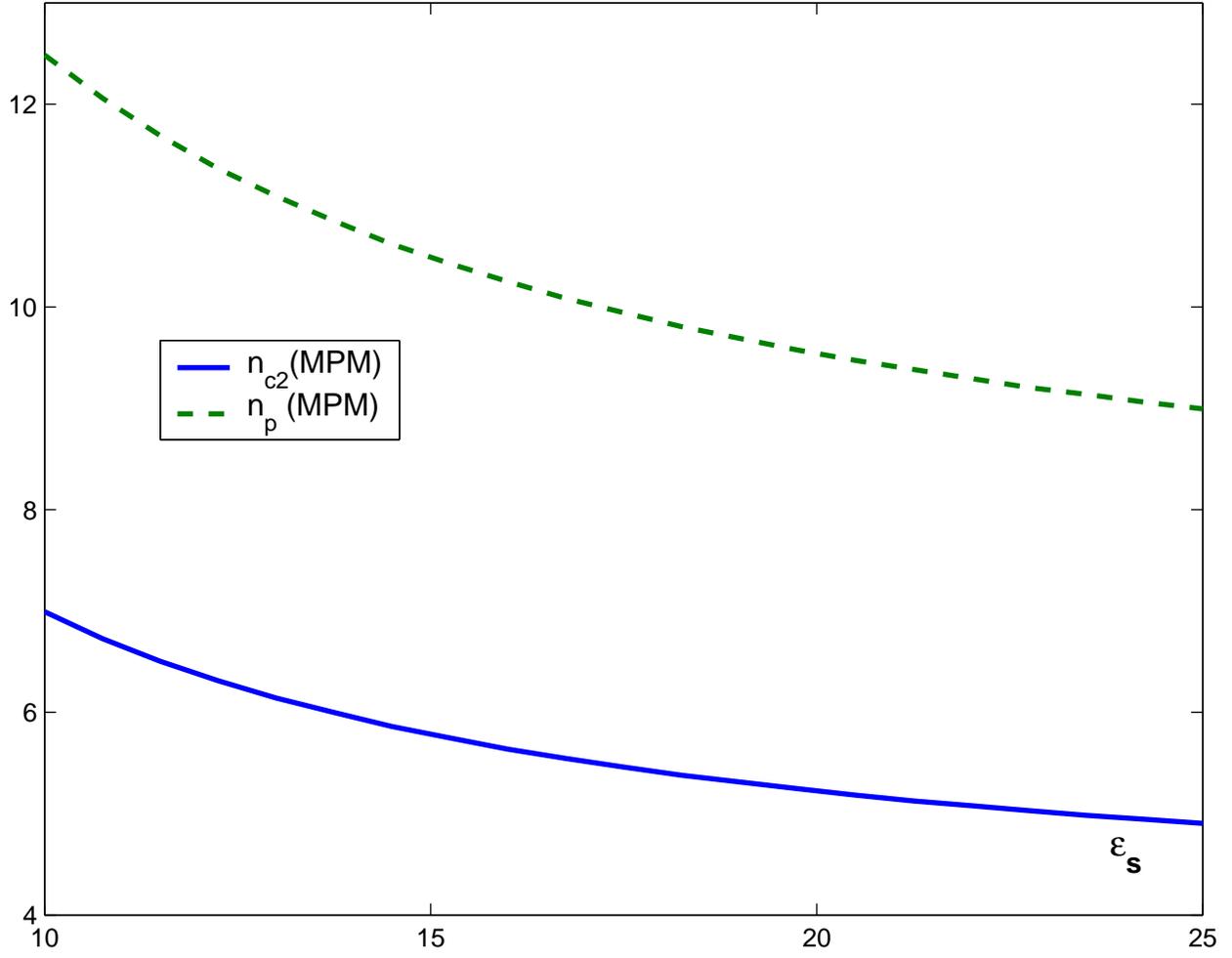}
\caption{Critical concentrations $n_{c2}$ and $n_p$ versus
low-frequency dielectric  constant $\epsilon_s$ for the metallic
phase of Na-NH$_3$  at $T=-70^0$C (model 1).} \label{fig3}
\end{figure}

\begin{figure}[tbp]
\includegraphics{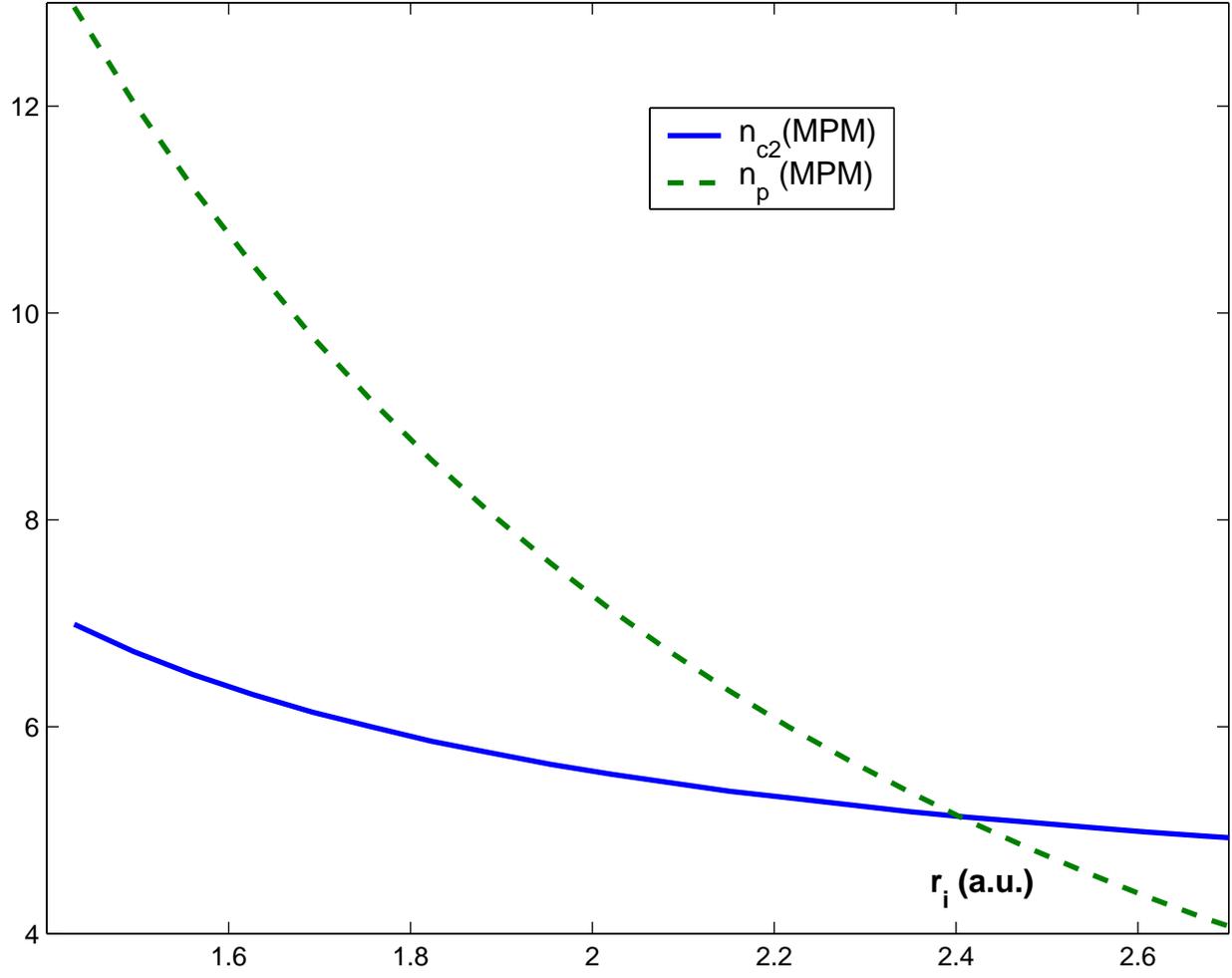}
\caption{Influence of the ion-core radius on the critical
concentrations $n_{c2}$ and $n_p$ in the case of the metallic phase
of Na-NH$_3$ at $T=-70^0$C (model 1).} \label{fig4}
\end{figure}

\begin{figure}[tbp]
\includegraphics{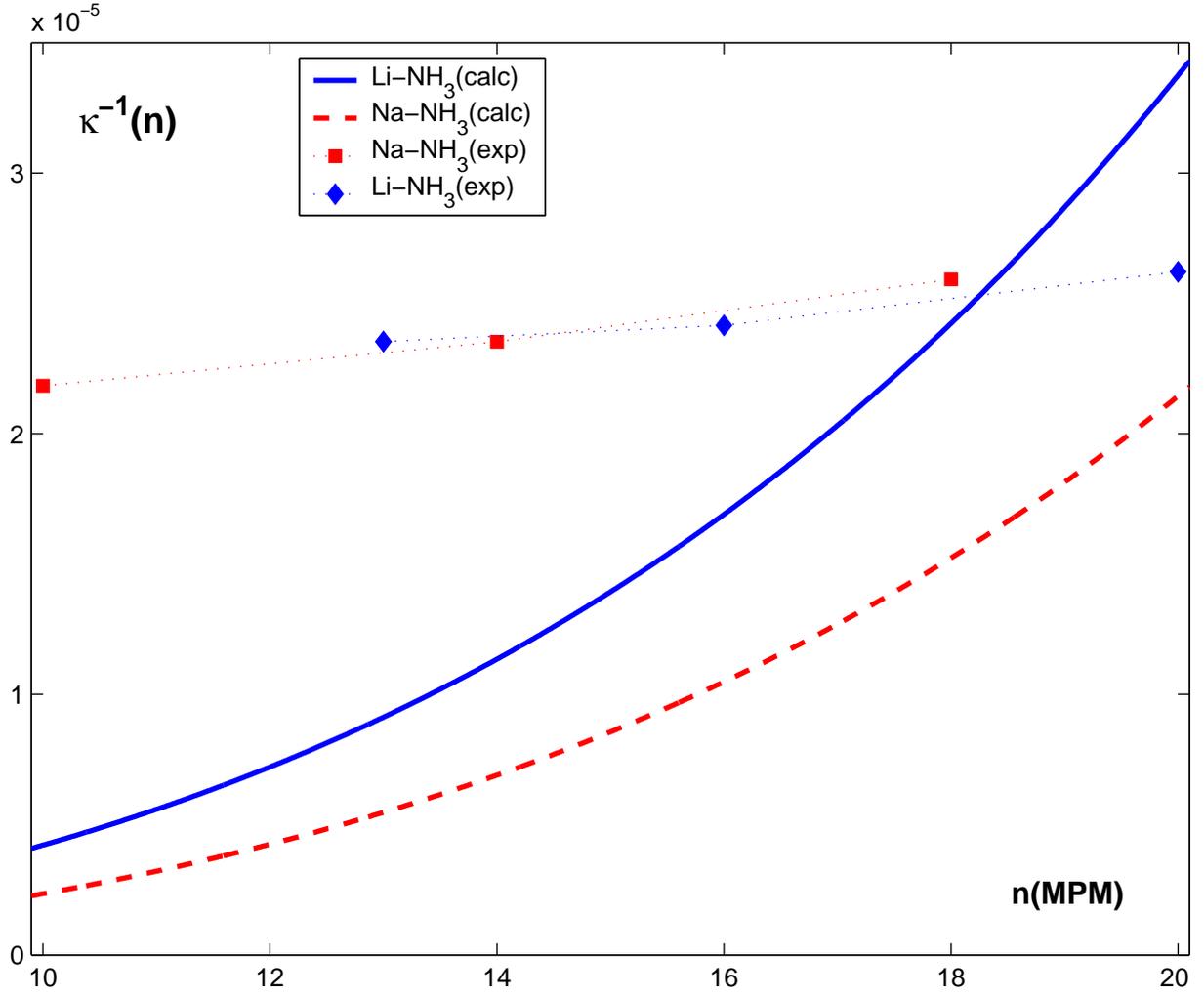}
\caption{The concentration dependence of the inverse compressibility
$\kappa^{-1}$ in MAS at $T=240^0$ K. The solid curve indicates our
calculations in the case of Li-NH$_3$ (model 2), and the diamonds
are the experimental data from \cite{burns2}, the dashed curve and
the squares are the same in the case of Na-NH$_3$ solutions.}
\label{fig5}
\end{figure}

\begin{figure}[tbp]
\includegraphics{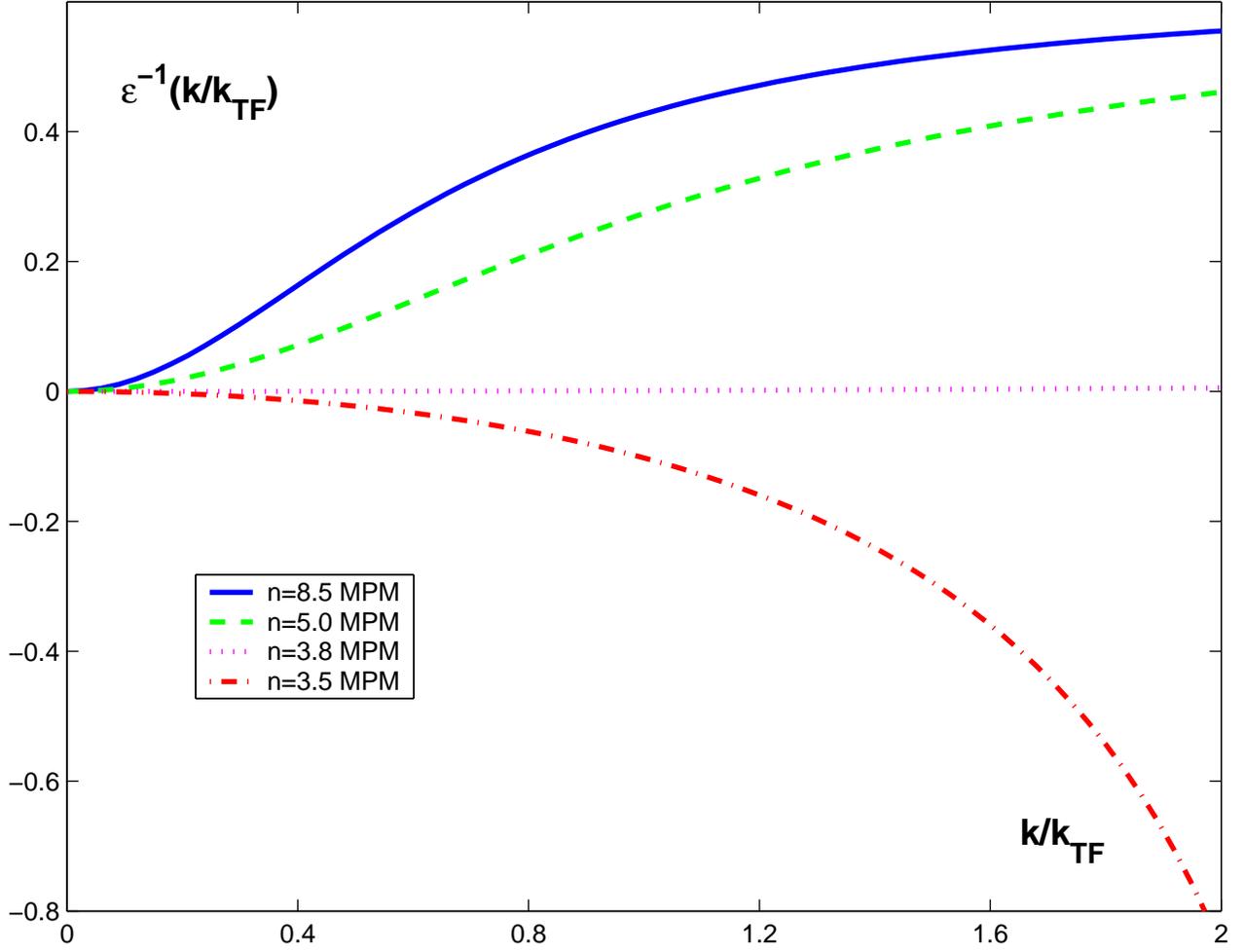}
\caption{The inverse static dielectric function
$\epsilon^{-1}(k,\omega=0)$ at various metal concentrations in
Li-NH$_3$ solutions at $T=240^0$ K (model 2). The values of metal
concentration $n$ are indicated at the corresponding lines.}
\label{fig6}
\end{figure}

\begin{figure}[tbp]
\includegraphics{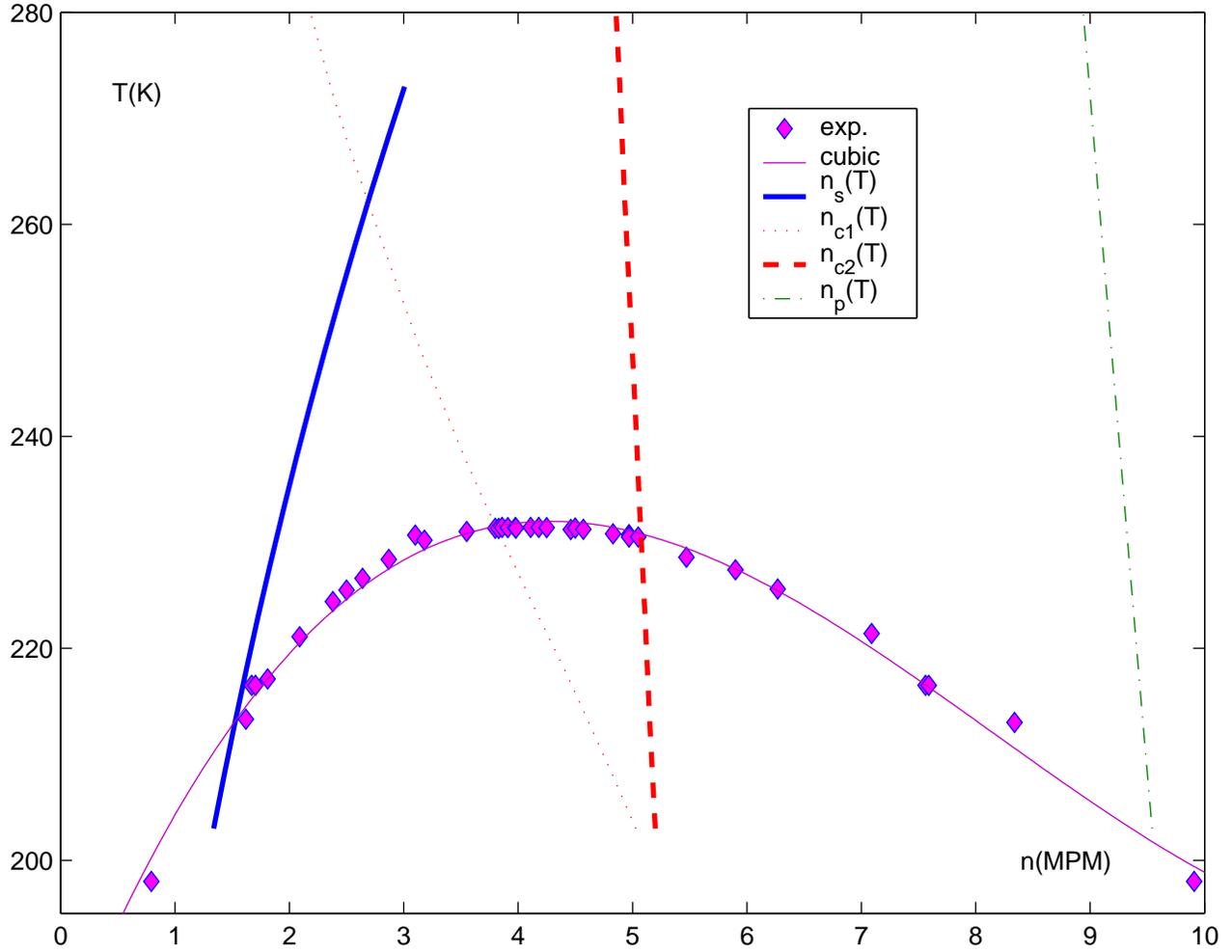}
\caption{The phase diagram of the Na-NH$_{3}$ solution and the lines
of instabilities calculated by  model 2. The experimental data
\cite{crauss,CHIEUX,rob} are indicated by symbols, and their cubic
interpolation is indicated by the thick curve. The dotted line
corresponds to the polarization catastrophe $n_{c1}$ considered as
the onset of metallization. The thin solid line is the low-density
spinodal $n_{s}$, the dashed one indicates the high-density spinodal
$n_{c2}$, and the dashed-dotted curve yields the zero-pressure line
of the pure metallic phase. The bottom of the figure corresponds to
the solidification of ammonia.}  \label{fig7}
\end{figure}

\begin{figure}[tbp]
\includegraphics{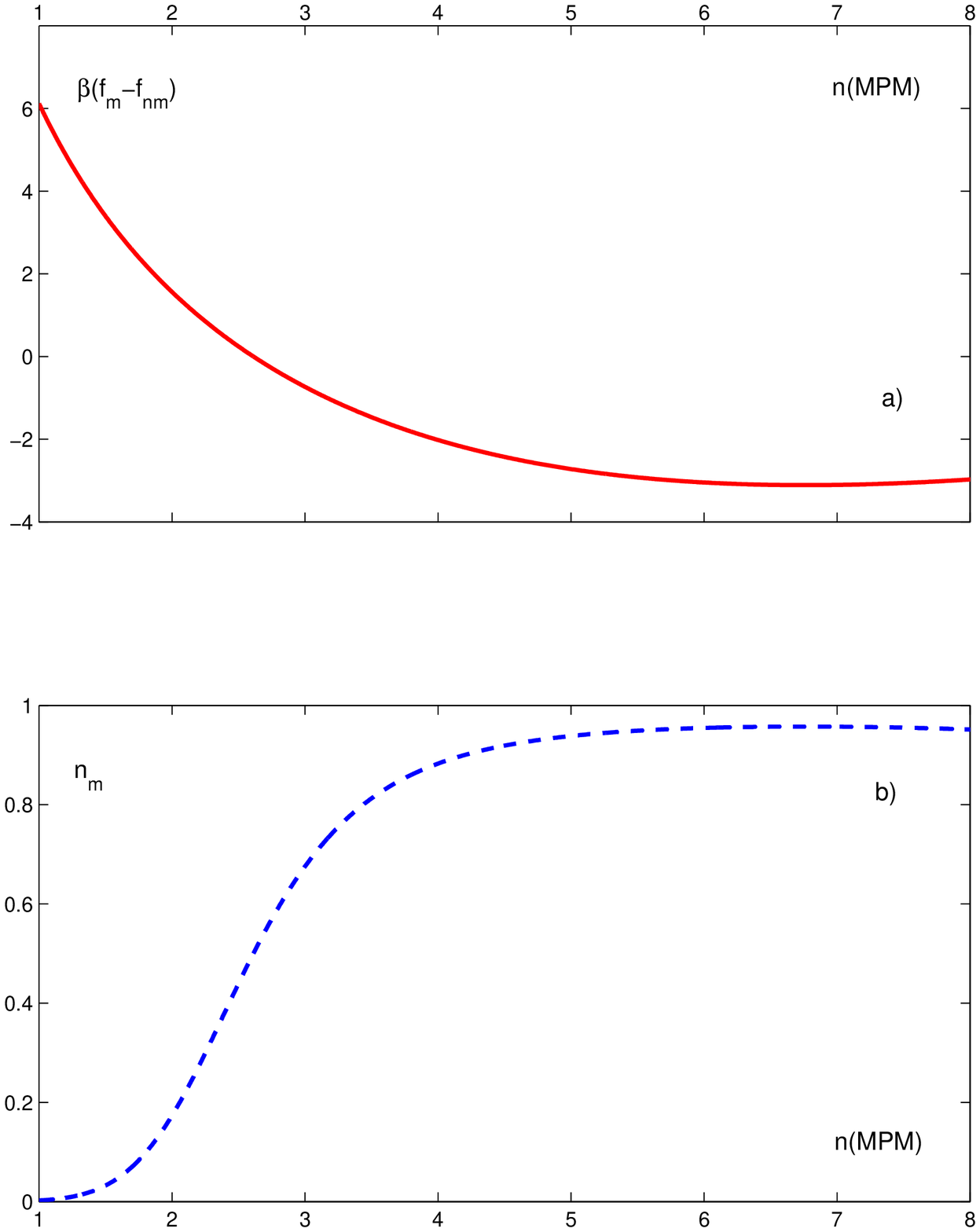}
\caption{The dimensionless difference $\beta(f_{m}-f_{nm})$ in the
free energies of metallic and nonmetallic phases (a) and the
relative fraction of electrons in delocalized states (b) versus the
metal concentration in the Na-NH$_3$ solution at $T=240^0$K, all
other parameters of the calculations correspond to Fig. 1.}
\label{fig8}
\end{figure}

\begin{figure}[tbp]
\caption{Schematic sketch of the influence of thermal fluctuations
on the concentration dependencies of the inverse compressibility
above (a) and below (b) critical temperature. The dashed-dotted line
corresponds to localized electrons, the dashed line represents the
inverse compressibility for delocalized electrons. The solid curves
correspond to the total inverse compressibility obtained as a sum of
the contributions of delocalized and localized electrons,
proportionally to their respective densities.} \label{fig9}
\end{figure}

\begin{figure}[tbp]
\includegraphics{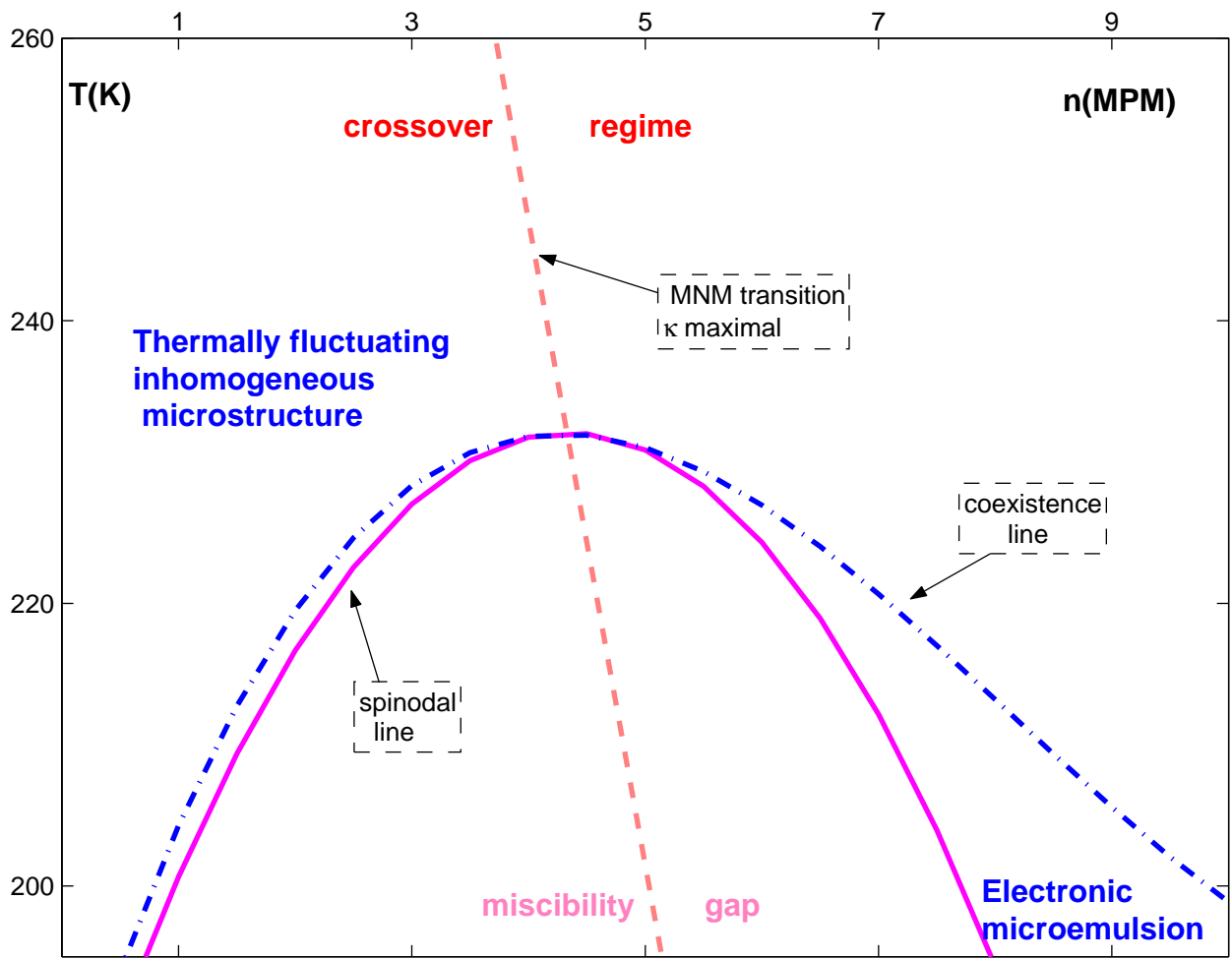}
\caption{Schematic sketch of the phase behavior of Na-NH$_3$
solutions. The dashed line corresponds to the MNM transition, which
is attributed to the line of the minimum inverse compressibility.
The solid curve is the spinodal line. The dashed-dotted curve
corresponds to the line of liquid-liquid coexistence.} \label{fig10}
\end{figure}


\begin{thebibliography}{99}
\bibitem{MPM}  Mole fraction is calculated as a ratio of moles of the metal
and that of the solution, 1(MPM) $\approx 2\cdot 10^{20}$ cm$^{-3}$.
\end{thebibliography}
\end{document}